\documentstyle[twocolumn,prl,aps]{revtex}
\begin{document}
\draft
\title{Stability of Relativistic Matter With Magnetic Fields}
\author{Elliott~H.~Lieb$^{1}$, Heinz Siedentop$^2$, and Jan Philip
Solovej$^3$} \address{$^1$Department of Physics, Jadwin Hall,
Princeton University, P.~O.~Box 708, Princeton, New Jersey 08544\\
$^2$Mathematik, Universit\"at Regensburg, D-93040 Regensburg, Germany\\
$^3$Department of Mathematics, Aarhus University, DK-8000 Aarhus C,
Denmark} 
\maketitle

\begin{abstract}
Stability of matter with Coulomb forces has been proved for
non-relativistic dynamics, including arbitrarily large magnetic fields,
and for relativistic dynamics without magnetic fields.  In both cases
stability requires that the fine structure constant $\alpha$ be not too
large. It was unclear what would happen for {\it both} relativistic
dynamics {\it and} magnetic fields, or even how to formulate the problem
clearly. We show that the use of the Dirac operator allows both effects,
provided the filled negative energy `sea' is defined properly.  The use
of the free Dirac operator to define the negative levels leads to
catastrophe for any  $\alpha$, but the use of the Dirac operator {\it
with} magnetic field leads to stability.
\end{abstract}

Since the work of Dyson and Lenard in 1967 \cite{DL} it has been well
understood that matter consisting of $N$ electrons and $K$ static (but
arbitrarily positioned), point nuclei, and governed by nonrelativistic
quantum mechanics, is stable. (See also \cite{LT} and see \cite{LR},
\cite{LG} for reviews.) This means that the ground state energy is
finite and is bounded below by a universal constant times the number
of particles, i.e., $E_0 \geq const. (N+K)$. If the nuclei are either
non-static or non-point the problem becomes easier.  The reason the
nuclei are taken to be static is that they are so massive compared to
the electron that if the nuclear mass (or the nuclear radius of
$10^{-13}$cm, for that matter) played any important role then
ordinary matter would have to look very different from what it does;
electron orbital radii would be many orders of magnitude smaller than
the Bohr radius.

We know that atoms, molecules and bulk matter exist and have ground
states that are well described by the Schr\"odinger equation with
infinitely massive, point nuclei.  This description needs corrections,
however, even at the low energies appropriate to atomic structure.
Effects due to special relativity and interactions of electrons with
(quantized or unquantized) magnetic fields are significant. Usually,
one appeals to quantum electrodynamics (QED) for these corrections,
but that theory, as usually formulated, is not known to exist outside
of perturbation theory.  More importantly, the effects mentioned are
often not perturbative, despite the fact that the fine structure
constant $\alpha = e^2/\hbar c \sim 1/137$ is small. It is well known,
for instance that the Dirac equation breaks down when $Z\alpha$ ($Z=$
nuclear charge), the quantity of greatest importance, is $1$.  The
quantity $Z\alpha$ is not at all small for Uranium and hence
relativistic and electrodynamic effects are not small. It is sometimes
said that some proper high energy theory (e.g., QCD or strings) will
straighten things out, but it is difficult to believe that we cannot
have a theory of chemistry without theoretical insertions in the GeV
range.
 
With this in mind, attempts have been made over the years to approach
a theory of matter by ever increasing generalizations of the many-body
Schr\"odinger equation, all the while maintaining the many-body aspect
and a non-perturbative, rigorous treatment.  One significant step was
the introduction of a `relativistic' kinetic energy into the
Schr\"odinger equation by replacing $p^2/2m$ by $ \sqrt{p^2
+m^2}-mc^2$ (with ${\bf p} = -i\hbar \nabla$ as usual) or, more
simply, by $p=|{\bf p}|$. We can also include a (non-quantized)
magnetic field ${\bf B}({\bf x})= {\rm curl}\, {\bf A}({\bf x})$
acting on the orbital motion, i.e., we use $|{\bf p}+{\bf A}({\bf
x})|$.  
The Hamiltonian (in appropriate units, in which all dimensional
parameters reside in $\alpha$ and $p=-i\nabla$) is then
\begin{equation}
H_{\rm rel} \equiv \sum_{i=1}^N |{\bf p}_i+{\bf A}({\bf x}_i)| +\alpha
V_{\rm c} \ ,
\label{eq:aj}
\end{equation}
where the Coulomb potential of the electrons (with coordinates ${\bf
x}_i$) and nuclei (with coordinates ${\bf R}_j$) is
\begin{eqnarray*}
V_{\rm c} &=& - Z\sum_{i=1}^N\sum_{j=1}^K |{\bf x}_i-{\bf R}_j|^{-1} +
\sum_{1 \le i<j \le N} |{\bf x}_i-{\bf x}_j|^{-1}\\ &&{}+ Z^2 \sum_{1
\le i<j \le K}|{\bf R}_i-{\bf R}_j|^{-1} \ .
\end{eqnarray*}
(It is purely for convenience that we take the nuclei to have the same
$Z$.)

For $A\equiv 0$, Conlon~\cite{C} showed that matter was stable in 
this theory
if $\alpha$ is `small enough'.  The constants were improved in
\cite{Fd} and finally in \cite{LiY} stability was proved all the way
up to the critical value appropriate to a `hydrogenic atom' (which is
$Z\alpha \leq 2/\pi$ in this theory \cite{H},\cite{K} 
instead of $Z\alpha \leq 1$ as in
the Dirac theory) {\it if and only if } $\alpha$ itself is less than
some critical value $\alpha_c$ (independent of $Z$). In \cite{LiY} it
was shown that $\alpha_c >1/94$.

For arbitrary, nonzero $A$ stability was shown to hold for
(\ref{eq:aj}) in \cite{LLSB} if $\alpha < \tilde \alpha$, where
\begin{equation}
1/\tilde\alpha \equiv ({\pi \over 2})Z + 2.80\ Z^{2/3} + 1.30\ ,
\label{eq:ak}
\end{equation}
which permits $Z\leq 59$ for $\alpha =1/137$. Eq. (\ref{eq:ak})
will be useful later.

Another important ingredient was the inclusion of arbitrarily large
magnetic fields and, at the same time the replacement of $p^2/2m$ by
the non-relativistic Pauli operator (in suitable units)
\begin{equation}
{\cal T}_{{\bf A}} \equiv [\mbox{\boldmath$\sigma$} \cdot ({\bf p} 
+ {\bf A}({\bf
x}))]^2= ({\bf p}+{\bf A}({\bf x}))^2+ \mbox{\boldmath$\sigma$} 
\cdot {\bf B} .\label{eq:aa}
\end{equation}

This field, ${\mathbf A}({\bf x})$, can be viewed in two ways: either as an
applied external field or as some sort of average (in the sense of
Hartree theory) of the field produced by all the electrons themselves.
The replacement of $p^2$ by $[({\bf p}+{\bf A}({\bf x})]^2$ is a
generalization that has long been known to cause {\it no} problems
with stability in the non-relativistic case. {\it The essential new
ingredient in the Pauli operator is the Zeeman term 
$\mbox{\boldmath$\sigma$} \cdot {\bf B}$, 
which results in the fact that there is no lower bound to the
energy, even for hydrogen \cite{AHS}.} (We have in mind here that we
make no restriction on the field ${\bf A}$, which is allowed to be
arbitrarily large.)  To remedy the instability due to arbitrarily
large fields we add the field energy
$$
H_{\rm field} \equiv [8\pi \alpha ]^{-1} \int B({\bf x})^2\, d^3x
$$
in the units we have been employing.

For a single atom it was shown a decade ago that the addition of
$H_{\rm field}$ stabilizes the system ($E_0$ is finite) if and only if
$Z\alpha^2$ is less than some comfortably large, but finite value
\cite{FLL}.
The `only if' part relies on a deep result in \cite{LY}.

Stability with arbitrary $N$ and $K$ and kinetic energy 
(\ref{eq:aa}) was first
solved by Fefferman (unpublished) for sufficiently small $\alpha$
and $Z$; see \cite{F} for an announcement. Shortly thereafter a
simple proof with good constants (e.g., we can take $Z=1050$ when
$\alpha =1/137$) was given \cite{LLSA} but, as in the `relativistic
case' mentioned above there is an $\alpha_c$ such that the system is
unstable whenever $\alpha > \alpha_c$, no matter how small Z may be.
Using the developments in \cite{LLSA}, it was shown in \cite{BFG} that
this non-relativistic theory remains stable when the magnetic field is
quantized, provided an ultraviolet cutoff is introduced.

Thus, two essential modifications of the Schr\"odinger energy
(replacing $p^2$ by $p$ or the inclusion of spin-magnetic field
interaction) require a bound on $\alpha$, in addition to a bound on
$Z\alpha$ or $Z\alpha^2$, for many-body stability. One might surmise
that the combination of the two would lead to further
difficulties---even, possibly, instability for all $\alpha$.  This is
not so, as we show in this letter. Indeed, our result is a bit
surprising for it reveals an unanticipated delicacy about the manner
of defining an electron in Dirac's theory.

The best way to include relativity and spin-field interaction is to
use the Dirac operator
$$
{\cal D}_{\bf A} \equiv {\bf p\hspace{-.16cm} \slash} 
+{{\mathbf A}\hskip-.2cm /} ({\bf x}) +\beta m .
$$
As usual,
${\mathbf O\hskip-.22cm/} \equiv \mbox{\boldmath$\alpha$} \cdot 
{\mathbf O}$, where $\alpha_i $ and
$\beta$ are the $4\times 4$ Dirac matrices.  Our many-body Hamiltonian
(with no pair production) is then, formally,
$$
H_{\rm Dirac}\equiv \sum_{i=1}^N {\cal D}_{\bf A}(i) +\alpha V_{\rm c}
+ H_{\rm field}.
$$

Unlike the usual (or the Pauli) kinetic energy operators, the Dirac
operator is not bounded below, so we use Dirac's prescription of
filling the negative energy `sea'. Concretely, this means that all the
electrons lie in the positive spectral subspace of the Dirac operator,
i.e., if $\Lambda^+$ denotes the projector onto the positive spectral
subspace for all the electrons then we allow only many-body functions
$\psi$ (which are antisymmetric functions of space and four component
spinors) that satisfy
$$
\Lambda^+ \psi = \psi .
$$
Another way to say this is that the ground state energy is given by
the infimum of the variational quantity (with respect to $\psi$ {\it
and} $A({\bf x})$)
$$
{\cal E}(\psi) \equiv \bigl\langle \psi| \Lambda^+ H_{\rm Dirac}
 \Lambda^+ |\psi \bigr\rangle \Big / \bigl\langle \Lambda^+ \psi | 
\Lambda^+ \psi \bigr\rangle .
$$

The idea of studying this problem 
goes back to  Brown and Ravenhall \cite{BR}, who
studied a single Helium atom with
${\mathbf A}={\mathbf 0}$ and $\Lambda_+$ being the 
positive spectral subspace for the free Dirac operator, 
${\cal D}_{\bf 0}$. This problem with a single general atom, 
i.e., $K=1$, but $N$ and $Z$ arbitrary,  has recently been shown 
stable, if $Z\alpha \leq 2/(\pi/2+2/\pi)$, and instable otherwise
\cite{EPS}.

While the Hamiltonian $H$ does not depend on the definition of the
positive spectral subspace, and it always includes the ${\mathbf A}$ 
field, the meaning of $\Lambda^+$, and hence $E_0$, does depend on 
the subspace.

There's the rub! Which Dirac operator should we use to define
$\Lambda^+$? There are at least two significant choices. One is the
free Dirac operator ${\cal D}_{\bf 0}= {\bf p\hskip-.16cm {\bf /}}
+\beta m$, which is the usual choice. The other is the Dirac operator
with the magnetic field ${\cal D}_{\bf A}$.  The latter seems more
important since an electron can never get rid of its own magnetic
field. Moreover, the choice ${\cal D}_{\bf 0}$, is {\it not gauge
invariant}, i.e., the multiplication of an electron wave function by a
spatially varying phase usually takes a positive energy function into
a mixture of positive and negative energy functions.  The second
choice is manifestly gauge invariant.

This issue is usually not clearly stated in field theory textbooks,
but whatever one might think about the appropriateness of either
definition the interesting and surprising fact is that ``stability''
of matter can settle the argument.  The following two theorems are the
main results in \cite{LSS}.  They show that one choice is valid and
the other is not.  
\bigskip

\noindent
{\bf THEOREM 1 (Stability with the magnetic Dirac operator):\/}
\smallskip

{\it If the electron wave function is restricted to lie in the
positive spectral subspace of the Dirac operator ${\cal D}_{{\bf A}}$
then $H_{\rm Dirac}$ is stable (i.e., $H_{\rm Dirac} >0$) provided
$\alpha $ and $Z$ are small enough. In particular, if we define
$\alpha_c$ to be the unique solution to the equation (with $\tilde
\alpha$ in (\ref{eq:ak}) and $L_{{1\over 2}, 3}$ defined below)
$$
1-(\alpha_c/\tilde \alpha)^2 = (16\pi L_{{1\over 2}, 3}
\alpha_c)^{2/3}
$$
then $\alpha \leq \alpha_c$ suffices for stability. In particular,
$Z=56$ and $\alpha=1/137$, or $Z=1$ and $\alpha =1/9$ are sufficient
for stability.}
\medskip

We shall outline the proof of this theorem here; details can be found
in \cite{LSS}.

\bigskip
\noindent
{\bf THEOREM 2 (Instability with the free Dirac operator):}
\smallskip
{\it If the electron wave function is restricted to lie in the
positive spectral subspace of the free Dirac operator ${\cal D}_0$
then the Hamiltonian $H_{\rm Dirac}$ is unstable. More precisely, for
any given values of $\alpha >0 $ and $Z>0$, there are (sufficiently
large) particle numbers $N$ and $K$ 
for which the infimum of ${\cal E}(\psi)$ is $-\infty$.}
\medskip

{\it Remark on the proof of Theorem 2:\/} This requires a complicated 
construction
of variational functions with arbitrarily negative energy.  It is
necessary to construct $N$-particle antisymmetric trial functions,
$\psi$, of space-spin {\it and also} corresponding magnetic potentials
${\bf A}({\bf x})$. See \cite{LSS}.
\bigskip

{\it Proof of Theorem 1:\/} Using (\ref{eq:aj}) and (\ref{eq:ak}) we
have that
$$
V_{\rm c} \geq -{1\over \tilde \alpha}\sum_{i=1}^N |{\bf p}_i+{\bf
A}({\bf x}_i)| \ .
$$
and hence ${\cal E}(\psi) -H_{\rm field}$ is bounded below by the sum
of the negative eigenvalues of the one-body operator
$$
h:= \Lambda^+\Bigl({\cal D}_{\bf A} - \kappa |{\bf p}+{\bf A}({\bf
x})| \Bigr)\Lambda^+,
$$
with $ \kappa \equiv\alpha / \tilde \alpha$. I.e., ${\cal E}(\psi)
-H_{\rm field} \geq - {\rm Tr} h_-$, where $h_- \equiv [|h|-h]/2$.

Next we note the BKS inequality \cite{BKS}: If $X>0,\ Y>0 $ are
self-adjoint operators, then
\begin{equation}
{\rm Tr} [X-Y ]_- \leq {\rm Tr} [ X^2-Y^2 ]_-^{1/2} .\label{eq:bks}
\end{equation}
We use (\ref{eq:bks}) with $X=\Lambda^+{\cal D}_{\bf A} \Lambda^+$ and
$Y=\Lambda^+ \kappa |{\bf p}+{\bf A}({\bf x})| \Lambda^+$. But we note
that ${\cal D}_{\bf A}^2 = {\cal T}_{\bf A} +m^2 > {\cal T}_{\bf A}
\geq ({\bf p} +{\bf A})^2 -B({\bf x})$ and we note that $X^2 =
\Lambda^+{\cal D}_{\bf A}^2 \Lambda^+$ (since ${\cal D}_{\bf A}$
commutes with $ \Lambda^+$), while $Y^2 \leq \kappa^2\Lambda^+ ({\bf
p}+{\bf A}({\bf x}))^2 \Lambda^+$. Thus we conclude that for every
choice of ${\bf A}$ and $\psi$

\begin{eqnarray}
{\cal E}(\psi) -H_{\rm field} &\geq& -{\rm Tr} \Bigl[ (1-\kappa^2)
({\bf p}+{\bf A}({\bf x}))^2 - B({\bf x}) \Bigr]_-^{1/2} \nonumber\\ &
\geq& -4 (1-\kappa^2)^{-3/2}L_{{1\over 2},3} \int B({\bf x})^2 \, d^3x
\label{eq:bg}
\end{eqnarray}
by the Lieb-Thirring inequality for the sum of the square roots of the
negative eigenvalues of a Schr\"odinger operator $({\bf p}+{\bf
A}({\bf x}))^2 + V({\bf x})$ with arbitrary ${\bf A}$ and arbitrary
potential $V$.  This is
\begin{equation}
{\rm Tr} \Bigl[\ ({\bf p}+{\bf A}({\bf x}))^2 +V({\bf
x})\Bigr]_-^{1/2} \leq L_{{1\over 2} , 3} \int V_-^2({\bf x}) \ d^3x
\label{eq:bh} \ .
\end{equation}
Note that in (\ref{eq:bg}) there is a factor of 4 on the right side
because the trace includes a trace over 4-dimensional spinors; in
(\ref{eq:bh}) there is no such trace, i.e., (\ref{eq:bh}) is for
spinless particles. It is known \cite{L} that $L_{{1\over 2},3} <
0.06003$ .

The factor 4 in (\ref{eq:bg}) can be reduced to 2, as one might expect
physically; details are in \cite{LSS}. Theorem 1 then follows easily from
(\ref{eq:bg}).

The authors thank the following organizations for their support: the
  Danish Science Foundation, the European Union, TMR grant FMRX-CT
  96-0001, the U.S. National Science Foundation, grant PHY95-13072A01,
  and NATO, grant CRG96011


\begin{thebibliography}{10}


\bibitem{DL}{F.J. Dyson and A. Lenard, J. Math. Phys. {\bf 8}, 423
(1967); {\bf 9}, 698 (1967). }

\bibitem{LT}{E.H. Lieb and W.E. Thirring, Phys. Rev. Lett. {\bf 35},
687 (1975); Errata {\bf 35}, 1116 (1975).}

\bibitem{LR}{E.H. Lieb, Rev. Mod. Phys. {\bf 48}, 553 (1976). }

\bibitem{LG}{E.H. Lieb, Bull. Amer. Math. Soc. {\bf 22}, 1 (1990).}

\bibitem{C}{J.G. Conlon, Commun. Math. Phys. {\bf 94}, 439 (1984).}

\bibitem{Fd}{C. Fefferman and R. de la Llave,
Rev. Math. Iberoamericana {\bf 2}, 119 (1986).}

\bibitem{LiY}{E.H. Lieb and H-T. Yau, Commun. Math. Phys. {\bf 118},
177 (1988).}

\bibitem{H}{I. Herbst, Commun. Math. Phys. {\bf 53}, 285 (1977).}

\bibitem{K}{T. Kato, {\it Perturbation Theory for Linear Operators} in
Grundl. der mathem. Wissen., {\bf 132}, Springer (1966).}

\bibitem{LLSB}{E.H. Lieb, M. Loss, and H. Siedentop, Helv. Phys. Acta,
{\bf 69}, 974 (1996).}

\bibitem{AHS}{J. Avron, I. Herbst, and B. Simon, Commun. Math. Phys.
{\bf 79}, 529 (1981).}

\bibitem{FLL}{J. Fr\"ohlich, E.H. Lieb and M. Loss,
Commun. Math. Phys.  {\bf 104}, 251 (1986); E.H. Lieb and M. Loss,
Commun. Math. Phys. {\bf 104}, 271 (1986).}

\bibitem{LY}{M. Loss and H-T. Yau, Commun. Math. Phys. {\bf 104}, 283
(1986).}

\bibitem{F}{C. Fefferman, Proc. Natl. Acad. Sci.  USA, {\bf 92},
5006 (1995).  }

\bibitem{LLSA}{E.H. Lieb, M. Loss, and J.P. Solovej, Phys. Rev. Lett.,
{\bf 75}, 985 (1995).}

\bibitem{BFG}{L. Bugliaro, J. Fr\"ohlich, and G.M. Graf,
Phys. Rev. Lett.  {\bf 77}, 3494 (1996).}

\bibitem{BR}{G.E. Brown and D.G. Ravenhall, Proc. Roy. Soc. London
{\bf A208}, 552 (1952).}

\bibitem{EPS}{D. Evans, P. Perry, and H. Siedentop,
Commun. Math. Phys. {\bf 178}, 733 (1996).}

\bibitem{LSS}{E.H. Lieb, H. Siedentop, and J.P. Solovej, {\it Stability
and Instability of Relativistic Electrons in Classical Electromagnetic
fields}, Jour. Stat. Phys., (accepted for publication).}
Preprint available at
http://xxx.lanl.gov/abs/cond-mat/9610195.


\bibitem{BKS}{M.S. Birman, L.S. Koplienko, and M.Z. Solomyak,
Sov. Math.  {\bf 19}, 1 (1975). Translation of Izvestija vyssich.}

\bibitem{L}{E.H. Lieb, Commun. Math. Phys. {\bf 92}, 473 (1984). }





\end{thebibliography}
\end{document}